\documentclass[12pt,a4paper]{article}

\usepackage[active]{srcltx} 
\usepackage{graphicx}
\usepackage{float}

\begin{document}
\textwidth=135mm
 \textheight=200mm
\begin{flushright}
    {\raggedright JLAB-THY-13-1686}
\end{flushright}
\begin{center}
{\bfseries \Large SPIN Effects, QCD, and Jefferson Laboratory with 12 GeV electrons
\footnote{{\small Summary of two plenary talks at SPIN 2012, Dubna, Russia}}}
\vskip 5mm
A. Prokudin$^{\dag}$
\vskip 5mm
{\small {\it $^\dag$ Thomas Jefferson National Accelerator Facility,\\
   Newport News, VA 23606, U.S.A. \\
prokudin@jlab.org} }
\\
\end{center}
\centerline{\bf Abstract}
QCD and Spin physics are playing important role in our understanding of hadron structure.
I will give a short overview of origin of hadron structure in QCD and highlight modern
understanding of the subject. Jefferson Laboratory is undergoing an upgrade that will
increase the energy of electron beam up to 12 GeV. JLab is one of the leading facilities 
in nuclear physics studies and once operational in 2015 JLab 12   will
be crucial for future of nuclear physics. I will briefly discuss 
future studies in four experimental halls of Jefferson Lab.
\begin{center}
 {\vskip -5mm PACs: 12.38.-t, 13.60.-r, 01.52.+r} 
\end{center}
 
\section{\label{sec:intro} Introduction}

With the advent of quark parton model and Bjorken scaling in 1960s the theoretical and 
experimental studies of the hadron structure became an important part of nuclear physics agenda
throughout the world.

Indeed by studying the proton we understand the underlying nature of Quantum Chromo Dynamics (QCD) -- the theory that
describes the hadron as bound system of quarks and gluons.  Asymptotic freedom of QCD allows one
to study the structure of the proton at small distances by varying, for example, the virtuality $Q^2$ of the
incident photon in Deep Inelastic Scattering. 

Protons are used as a discovery tool
in several facilities including Large Hadron Collider and precise knowledge of its structure becomes 
an essential ingredient of the discovery potential of such facilities. 

A number of experimental facilities study hadron structure. In particular experimental studies including
spin degrees of freedom are important. HERMES (DESY), COMPASS (CERN), RHIC (BNL), JLAB  pioneered these studies.
Fragmentation of quarks into colorless hadrons are being studied at  BELLE (KEK) and BaBar (SLAC).

Jefferson Lab is accomplishing the 12 GeV upgrade project \cite{Dudek:2012vr} which is due to be operational in 
2015 and will enable us to look with an 
unprecedented precision at the nucleon structure in the region where valence quarks are dominant in
 nucleon's waive function. Such precision is needed for better understanding of the nature of the nucleon
as a many body relativistic system in terms of internal dynamics.  

\begin{figure}[H]
\centering \includegraphics[width = 0.45\textwidth]{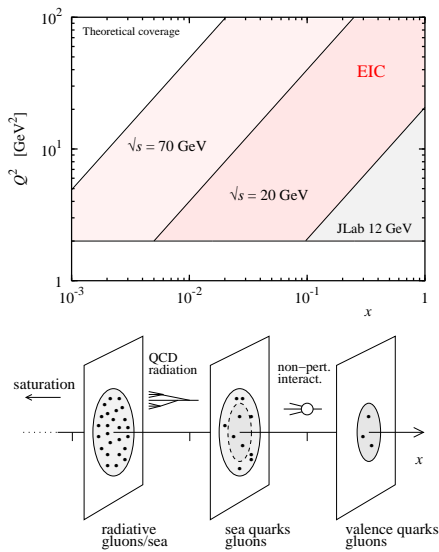}
\caption{(Top) Kinematic coverage in $x$ and $Q^2$ in ep scattering
experiments with JLab 12 GeV and a medium-energy EIC of
CM energy $\sqrt{s}$ = 20 and 70 GeV. The minimum momentum
transfer here was chosen as $Q^2_{min}$ = 2 GeV$^2$. (Bottom) Components 
of the nucleon wave function probed in scattering experiments
at different $x$.\label{fig1}}
\end{figure}

Looking forward in future one would like to study the dynamical origin of quarks ad gluons in the region where sea quarks
and gluons start dominating nucleon's waive function. This can be achieved by constructing a new facility -- polarized Electron Ion Collider \cite{Dudek:2012vr,Accardi:2011mz,Abeyratne:2012ah}
 or EIC with variable center-of-mass energy $\sqrt{s}$ $\sim$ 20 --70 GeV and luminosity $\sim$ $10^{34}$ cm$^{-2}$ s$^{-1}$
 that  would be uniquely suited to address several outstanding
questions of Quantum Chromodynamics (QCD) and the microscopic structure of hadrons and nuclei. In Fig.~\ref{fig1}
\footnote{The plot is from Ref.~\cite{Accardi:2011mz}. See Ref.~\cite{Accardi:2011mz} for details on nuclear physics 
opportunities at a medium-energy EIC.} kinematical ranges of JLab and EIC are
compared as functions of Bjorken-x and $Q^2$.

Spin and polarization measurements have been playing a crucial role in our understanding of nucleon's properties throughout many decades.
Since famous ``Spin crisis'' \cite{Aubert:1985fx,Leader:1988vd} of 1980's we learned that quark spins do not account for the
full spin of the nucleon. Given the later observation that the contribution of the gluon spin to that of the nucleon
could be rather small \cite{deFlorian:2011ia} one concludes that a static picture of the nucleon with quarks 
in $s$-states does not account for the complexity of the parton dynamics. Orbital motion of quarks and gluons must play an important role
in our understanding of the nucleon's structure. 

In recent years the description of the nucleon's spin and momentum 
structure given in terms of partonic sub-structure has led to rapid development of QCD theory. In hard semi-inclusive processes
 involving non-collinear dynamics these structures are described 
by Transverse Momentum Parton distributions and fragmentation functions (TMD-PDFs and TMD-FFs, or jointly TMDs). TMDs 
depend both on Bjorken--$x$ and transverse motion of partons $\bf k_T$ thus making them sensitive to Orbital Angular
Momentum of quarks and gluons.
The transverse degrees of freedom also play a crucial role in high energy collider experiments 
through so called Efremov-Teryaev-Qiu-Sterman matrix elements \cite{Efremov:1981sh,Efremov:1984ip,Qiu:1991pp} i.e. multi-parton correlations. 

In more exclusive processes such as Deep Virtual Compton Scattering or Exclusive Vector Meson Electro production
one encounters so-called Generalized Parton Distributions (GPDs) that, by Fourier transform over
transferred momentum $t$, depend additionally to the usual Bjorken--$x$ 
on the position of partons in coordinate space.

\begin{figure}[H]
\centering \includegraphics[width = 0.53\textwidth]{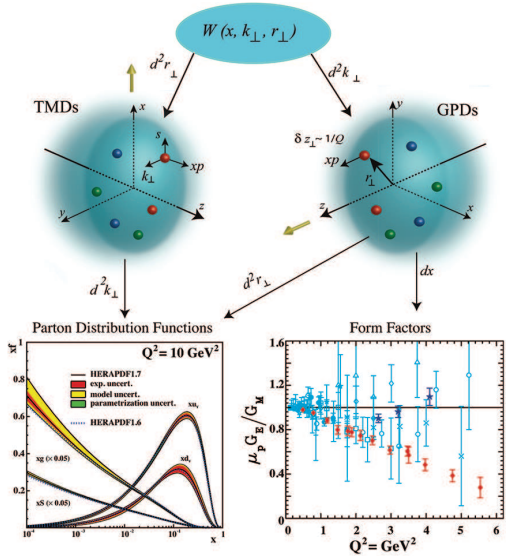}
\caption{Wigner distribution and relation to Generalized Parton Distributions and Transverse Momentum Dependent Distributions. Parton distributions 
and form factors can be related to GPDs and TMDs. \label{fig2}}
\end{figure}

There is a general belief that QCD is the underlying theory that describes nucleon structure by quark and gluon 
degrees of freedom, yet 
we lack a detailed understanding of these objects from first principles. Nevertheless, a new framework has emerged 
in the past ten years which is suitable for a comprehensive and quantitative approach to the description of nucleon structure 
\cite{Ji:2003ak,Belitsky:2003nz,Belitsky:2005qn}. 
In this framework our knowledge of nucleon structure is encoded in the Wigner distributions of the constituents, 
a quantum mechanical concept, introduced in 1932 \cite{Wigner:1932eb}. 
From the Wigner distributions, see Fig.~\ref{fig2}\footnote{The plot is from Ref.~\cite{Dudek:2012vr}}, a natural interpretation of measured observables is provided through the construction of 
its integrated ``slices'' or projections which are in fact Generalized Parton Distributions and Transverse Momentum Dependent distributions.

\section{\label{sec:qcd} QCD evolution and SPIN effects}

The nucleon in QCD represents a dynamical system of
fascinating complexity. In the rest frame it may be viewed
as an ensemble of interacting color fields, coupled in an
intricate way to the vacuum fluctuations that govern the
effective dynamics at distances $\sim$ 1 fm.   A
complementary description emerges when one considers a
nucleon that moves fast, with a momentum much larger
than that of the typical vacuum fluctuations. In this limit
the nucleon's color fields can be projected on elementary
quanta with point-particle characteristics (partons), and
the nucleon becomes a many-body system of quarks and
gluons. As such it can be described by a wave function, in
much the same way as many-body systems in nuclear or
condensed matter physics. In contrast to these
non-relativistic systems, in QCD the number of point-like
constituents is not fixed, as they constantly undergo creation/
annihilation processes mediated by QCD interactions,
reflecting the essentially relativistic nature of the
dynamics.

Accordingly the QCD evolution that governs content of the nucleon 
is interpreted differently in different frames. 

If one considers evolution of parton
densities with energy than the appropriate frame is so-called dipole frame
in which virtual photon fluctuates into a color dipole (quark--antiquark pair) and this dipole
interacts with target nucleon. Corresponding evolution is governed by Balitsky-Fadin-Kuraev-Lipatov (BFKL)
evolution equation \cite{Kuraev:1977fs,Balitsky:1978ic}. The non linear regime of this evolution is described via Balitsky equation 
\cite{Balitsky:1995ub}
Balitsky-Kovchegov equation in large $N_c$ limit 
(BK) \cite{Balitsky:1995ub,Kovchegov:1999yj,Kovchegov:1999ua} and JIMWLK evolution equations 
\cite{JalilianMarian:1997gr,Iancu:2000hn,Ferreiro:2001qy}. Subsequently the system will pass from dilute to dense 
regime of QCD and to predicted but yet to be observed regime of saturation of gluon densities. Geometrical
scaling of structure functions at low-$x$ observed at HERA (DESY) \cite{Schildknecht:2000zt} is an indication of this regime to
take place. Note that the resolution scale that is defined
by the virtuality of the photon $Q^2$ is fixed in this case.

\begin{figure}[t]
\centering \includegraphics[width = 0.4\textwidth]{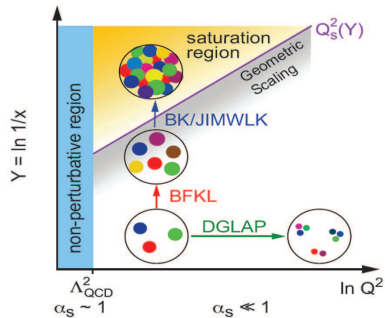}
\caption{Evolution of parton densities can be considered either in energy/rapidity $y$ (BFKL, BK and JIMWLK equations) or in virtuality of the photon $Q^2$ 
(DGLAP equation). The system will go from dilute towards dense regime and undergo transition to saturation region which is characterized by saturation scale $Q^2_s(y)$.
\label{fig3}}
\end{figure}

DGLAP equation describes the evolution of densities as function of $Q^2$ at given energy scale or rapidity $y$. Infinite 
Momentum Frame (the frame in which the target nucleon moves with infinite momentum and thus along light-cone) is suitable for interpretation in this case.
Fluctuations of incident photon into quark-antiquark pairs are suppressed and the photon probes ``frozen'' partonic states inside
of the nucleon.
Gluon radiation in the available phase space produces multiple quark, antiquark and gluon states that are responsible for the
growth of parton densities in low-$x$ region. Note that the resolution scale $Q^2$ increases and thus the distance at which the
states are probed and the ``effective size'' of partons diminishes. See Fig.~\ref{fig3} for
representation of different evolutions.

Evolution of Transverse Momentum Dependent distributions is an emerging subject of nuclear theory. The details of
TMD factorization were derived in Ref.~\cite{CollinsBook} and successfully implemented in Refs.~\cite{Aybat:2011zv,Aybat:2011ge}.
It was demonstrated that TMD evolution \cite{Aybat:2011ta,Anselmino:2012aa} appropriately takes into account the behavior of experimental data. One of the particularities of the TMD evolution consists in fact that
unlike usual collinear distributions where only collinear singularities are present, TMDs exhibit rapidity divergences 
along with collinear ones. Thus evolution is more intricate and describes not only how the form of distribution changes in 
terms of Bjorken-$x$ but also how the width is changed in momentum space $\bf k_T$. It was shown in Ref.~\cite{CollinsBook} that TMD formalism in fact corresponds
to well known Collins-Soper-Sterman (CSS) resummation \cite{Collins:1981uw,Collins:1984kg}.

Evolution of twist-3 matrix elements was also recently worked out in Refs~\cite{Kang:2008ey,Zhou:2008mz,Vogelsang:2009pj,Braun:2009mi} and the obtained result by three groups employing different methods
agree with each other \cite{Kang:2012em}. The CSS resummation was also applied to spin dependent quantities in Ref.~\cite{Kang:2011mr}. Along
with advances in TMD evolution implementation these results will lead to complete NLO knowledge of TMDs and twist-3 matrix elements which are sources
of spin asymmetries observed in different experiments in SIDIS, DY, and $e^+e^-$ annihilation.

Many new formulations of TMD factorization \cite{Cherednikov:2009wk} and in particular in the framework of Soft Collinear Effective Theory (SCET) have emerged 
recently \cite{GarciaEchevarria:2011rb,Echevarria:2012pw,Chiu:2012ir}. General relations between those different formulations and comparison of resulting evolution equations will be particularly 
interesting in future. 
 
\section{\label{sec:puzzle} Puzzles of SPIN}
The ``Spin crisis'' \cite{Aubert:1985fx,Leader:1988vd} of 1980's was not the last one to challenge our theoretical understanding 
of hadron structure and QCD. There existed a simple and intuitive prediction \cite{Kane:1978nd} for the so-called $A_N$ asymmetry
in $pp^\uparrow \leftarrow \pi X$ to be negligible. Famous measurement of FNAL-E704 \cite{Adams:1991ru} proved this prediction to be wrong.
Not only the asymmetry was large at relatively low energy $\sqrt{s} = 19.4$ GeV \cite{Adams:1991ru} , but it remained so at much higher energies at RHIC up to $\sqrt{s} = 200$ GeV  \cite{Lee:2007zzh,:2008qb}.

For processes such as single inclusive hadron production in proton-proton collisions, 
$p^\uparrow p\to hX$,
which exhibits only one characteristic hard scale, the transverse momentum $P_{h\perp}^2 \gg \Lambda_{\rm QCD}^2$ of the produced hadron, one could describe the SSAs in terms of twist-three quark-gluon correlation functions \cite{Efremov:1981sh,Efremov:1984ip,Qiu:1991pp,Qiu:1998ia,Koike:2009ge,Kang:2010zzb}. 
One of the well-known examples is the so-called Efremov-Teryaev-Qiu-Sterman (ETQS) function. 
Phenomenological extractions were performed in different papers \cite{Kouvaris:2006zy,Kanazawa:2011bg}. 

On the other hand, for processes such as Semi-Inclusive Deep Inelastic Scattering (SIDIS) which possesses two characteristic scales, photon's virtuality $Q$ and $P_{h\perp}$ of the produced hadron, one can use a TMD factorization formalism \cite{Ji:2004wu,Ji:2004xq, CollinsBook} 
 in the region $ \Lambda_{\rm QCD}^2 < P_{h\perp}^2 \ll Q^2 $ and describe 
asymmetries with TMD functions. Extractions of TMDs have been performed 
using experimental data at fixed scales 
 ~\cite{Efremov:2004tp,Vogelsang:2005cs,Anselmino:2005nn,Arnold:2008ap,Anselmino:2008sga,Anselmino:2008jk,Anselmino:2007fs}.

These two formalisms are closely related to each other, and have been shown to be equivalent in the overlap region where both can apply \cite{Ji:2006ub,Koike:2007dg,Bacchetta:2008xw}.

Recently it has been found that there exists ``sign puzzle'' or ``sign mismatch'' between these two mechanisms, \cite{Kang:2011hk}. Yet another
puzzle to challenge our understanding of QCD. Some preliminary explanations are already available \cite{Kang:2012xf}, however in order to achieve 
the complete coherent picture we will have to work for more years to come.

\section{Jefferson Lab with 12 GeV electrons}

The continuous Electron Beam Accelerator Facility (CEBAF) of Jefferson Lab is being upgraded and will
provide electron beam of 11 GeV to three experimental Halls A, B, and C and 12 GeV electron beam to HALL D. CEBAF will also maintain capability of providing lower energy beam to the Halls.

\begin{figure}[H]
\centering \includegraphics[width = 0.45\textwidth]{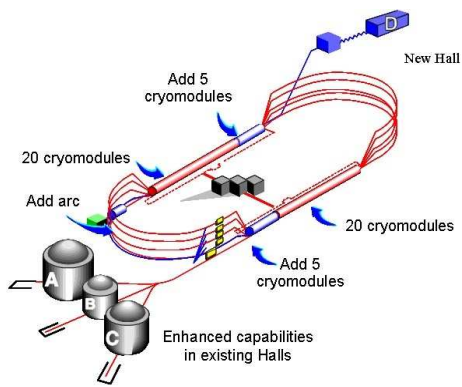}
\caption{CEBAF of Jefferson Lab and four experimental Halls. \label{jlab}}
\end{figure}

Jefferson Lab itself is a multi purpose laboratory for nuclear studies. Its scientific activity spans from
material studies, lasers, medical imaging, accelerator research and development to a vast fundamental
experimental and theoretical research in nuclear physics and searches beyond standard model.

6 GeV scientific program of Jefferson Lab successfully finished in 2012. Upgrade is designed to build on existing
facility: vast majority of accelerator and
experimental equipment have continued use. The completion of the 12
GeV Upgrade of CEBAF
was ranked the highest
priority in the 2007 NSAC
Long Range Plan. The scope of the project includes doubling the accelerator beam energy,
construction of a new experimental Hall (D) and beamline, an
upgrading to existing experimental Halls (A,B,C).

The main goals of Jefferson Lab experimental program are
\begin{itemize}
\item The physical origins of quark confinement (meson and baryon spectroscopy)
\item The spin and flavor structure of the proton and neutron (PDFs, GPDs, TMDs)
\item The quark structure of nuclei
\item Probe potential new physics through high precision tests of the Standard Model
\end{itemize}

In order to define the scientific program Jefferson Lab Program Advisory Committee gathered 8 times 
in the period 2006 -- 2011 and as a result 52 experiments were approved and 15 experiments we
conditionally approved. White paper \cite{Dudek:2012vr} was submitted for NSAC subcommittee.

Hall D will be exploring origin of confinement by
studying exotic mesons. In order to study mesonic system photon beam of energy up to 9 GeV
will be produced. GlueX experiment being constructed in Hall D will reach the mass range up to $3.5$ GeV and 
will offer insight into the role of gluon self interactions and the nature of confinement.
Detailed spectroscopic information from experiment, coupled with the guidance of new Lattice QCD results, offers an exciting and unique opportunity to explore mechanisms of confinement.

HERMES and COMPASS, together with the 6 GeV Jefferson Lab have demonstrated the feasibility of studying Transverse Momentum Dependent distributions (TMDs) as well as Deeply Virtual Compton Scattering (DVCS) measurements that offer access to Generalized Parton Distributions (GPDs). The extended kinematic range and new experimental hardware associated with the Jefferson Lab 12 GeV. Upgrade will provide access to these fundamental underlying distributions and reveal new aspects of nucleon structure. It is quite possible that much of the remaining nucleon spin will be found in the orbital motion of the valence quarks. HALLS A, B, and C have 18 approved experiments dedicated to studies of TMDs and GPDs. HALL B CLAS detector will have hermetic design which is important for exclusive reaction measurements. Future data from the corresponding experiments in Hall B with CLAS 12, in Hall A with Super-BigBite and with SoLID complemented with precision SIDIS experiments in Hall C will allow a far more precise determination of TMDs, GPDs and ordinary PDFs to a much greater precision if compared to modern knowledge of these distributions.

The electric and magnetic form factors of the nucleon describe the distribution of charge and currents, and are probed in elastic electron-nucleon scattering. JLab 12 will continue studying form factors and reach much higher values of $Q^2$ up to 10 -- 11 GeV$^2$.

11 experiments in HALLS A, B, and C are dedicated to studies of hadrons and cold nuclear matter. One of the outstanding questions is whether the nuclear medium alters the structure of bound nucleons and, if it does, how?

It is believed that Standard Model as a theory of fundamental interactions is incomplete. Thus it is important to pursue precision tests and searches
beyond Standard Model. JLab 12 with its high luminosity and accuracy is certainly one of the key payers in this search. A very precise SM prediction of running of  $\sin^2\theta_W$, where $\theta_W$ is the weak mixing angle, allows for a precision test of Standard Model. High luminosity of JLab 12 up to $10^{38}$ cm$^{-2}$ s$^{-1}$ is certainly one the key ingredients for successful high precision measurement. 

The Qweak experiment has completed data collection to measure $A_{PV}$ in elastic electron-proton scattering at low $Q^2$ 0.021 GeV$^2$ in Hall C
\cite{qweak}. The weak charge of the proton $Q^p_
W = 1 - 4 \sin^2\theta_W$ is suppressed, which allows for search of beyond standard model contributions. This suppression and the expected precision
of the Qweak measurement of $Q^{p}_
{W}$ of 4\%, gives a sensitivity to new parity-violating physics up to 2 TeV. Parity violation experiments require polarized electrons, which are routinely produced already at CEBAF, and many of the electroproduction experiments planned, e.g. DVCS experiments also require polarized electrons. MOLLER experiment and SoLID will continue PV measurements at JLab 12.

Heavy photons, called A$'$s, are new hypothesized massive vector bosons that have a small coupling to electrically charged matter, including electrons. The existence of A$'$ can explain discrepancy between measured and predicted value of anomalous magnetic moment of the muon \cite{Pospelov:2008zw}. Moreover signals of astrophysical positron excess \cite{Adriani:2008zr} suggest
 a massive neutral vector boson A$'$
with low mass ($M_{A'} <$ 1 GeV ). APEX (Hall A), HPS (Hall B), and Dark Light (FEL) will search for A$'$ in MeV-GeV mass range.

Concluding we might say that JLab 12 will provide decades of extremely interesting reseach and measurements in nuclear
physics and beyond.
In no way the information presented here accounts completely for all plans of JLab 12, interested reader is referred to the White paper \cite{Dudek:2012vr} for more information.

\section*{Acknowledgement}
The author thanks Hugh Montgomery for careful reading of the manuscript and Anatoly Radyushkin for useful discussions.
Authored by a Jefferson Science Associate, LLC under U.S. DOE Contract 
No. DE-AC05-06OR23177.  


\end{document}